\documentclass[structabstract]{aa}  
\usepackage{graphicx}
\usepackage{txfonts}
\usepackage{amssymb}
\usepackage{longtable}
\usepackage{supertabular}
\usepackage{natbib}
\begin{document}
   \title{Deep GMRT radio observations and a multi-wavelength study of the region around HESS J1858+020}


   \author{J.M. Paredes\inst{1}
          \and
          C. H. Ishwara-Chandra\inst{2}
          \and
          V. Bosch-Ramon\inst{1}
          \and
          V. Zabalza\inst{3}
          \and
          K. Iwasawa\inst{4}
          \and
           M. Rib\' o \inst{1}
                    }

   \institute{Departament d'Astronomia i Meteorologia, Institut de Ci\`encies del Cosmos, Universitat de Barcelona, IEEC-UB, Mart\'{\i} i Franqu\`es 1, E-08028, Barcelona, Spain\\
              \email{jmparedes@ub.edu; vbosch@am.ub.es; mribo@am.ub.es}
         \and
             National Centre for Radio Astrophysics, TIFR,  Post Bag 3, Ganeshkhind, Pune -  411007, India\\
             \email{ishwar@ncra.tifr.res.in}
          \and
          Max-Planck Institut f\"ur Kernphysik, Postfach 103980, D-69029 Heidelberg, Germany\\
              \email{Victor.Zabalza@mpi-hd.mpg.de}
              \and
            ICREA, Institut de Ci\`encies del Cosmos, Universitat de Barcelona, IEEC-UB, Mart\'{\i} i Franqu\`es 1, E-08028, Barcelona, Spain\\
              \email{kazushi.iwasawa@icc.ub.edu} 
                         }

   \date{Received ; accepted }

 
  \abstract
   {There are a number of very high energy sources in the Galaxy that remain unidentified. Multi-wavelength and variability studies, and catalogue searches, are powerful tools to identify the physical counterpart, given the uncertainty in the source location and extension.}
   {This work carries out a thorough multi-wavelength study of  the  unidentified, very high energy source HESS~J1858+020 and its environs.}
   {We have performed Giant Metrewave Radio Telescope observations at 610 MHz and 1.4 GHz to obtain a deep, low-frequency radio image of the region surrounding HESS~J1858+020. We analysed archival radio, infrared, and X-ray data as well. This observational information, combined with molecular data, catalogue sources, and a nearby $Fermi$  gamma-ray detection of unidentified origin, are combined to explore possible counterparts to the very high energy source.}
   {We provide with a deep radio image of a supernova remnant that might be related to the GeV and TeV emission in the region. We confirm the presence of an \ion{H}{ii} region next to the supernova remnant and coincident with molecular emission. A potential region of star formation is also identified. We identify several radio and X-ray sources in the surroundings. Some of these sources are known planetary nebulae, whereas others may be non-thermal extended emitters and embedded young stellar objects. Three old, background Galactic pulsars also neighbour HESS~J1858+020 along the line of sight.}
   {The region surrounding HESS~J1858+020  is rich in molecular structures and non-thermal objects that may potentially be linked to this unidentified very high energy source. In particular, a supernova remnant interacting with nearby molecular clouds may be a good candidate, but a star forming region, or a non-thermal radio source of yet unclear nature, may also be behind the gamma-ray source. The neighbouring pulsars, despite being old and distant, cannot be discarded as candidates. Further observational studies are needed, however, to narrow the search for a counterpart to the HESS source.}


   \maketitle

\section{Introduction}

The High Energy Stereoscopic System (H.E.S.S.) survey of the inner Galaxy \citep{2006ApJ...636..777A} has revealed new sources that remain
unidentified because they have not been associated with a particular object from which very high energy (VHE: $E > 100 $ GeV)
emission is expected. A study of the environment of eight of these
unidentified extended TeV sources with high detection significance revealed no
plausible counterpart  \citep{aharonian08}. 
These eight sources are
extended, with angular sizes ranging from 3 to 18 arcmin. Their spectra in the
TeV energy range follow a power-law with a differential spectral index in the
range 2.1 to 2.5. The source sizes and spectra are similar to previously identified
Galactic VHE sources (e.g., PWNe). 
In some of these sources, a plausible  counterpart has been found (HESS~J1731$-$347, \citealt{2010ApJ...712..790T}; HESS~J1841$-$055,  \citealt{2012ApJ...755L..20P}) or has been described as ancient PWNe (e.g., HESS~J1427$-$608, \citealt{2012AIPC.1505..349T}) in the last few years, but other sources, such as  HESS~J1858+020, still lack a plausible counterpart at low energies.  

The source HESS~J1858+020 is one of the weakest gamma-ray sources of the \cite{aharonian08} sample and it shows a slight extension along its major axis, in the north-south direction. Close to it lies HESS~J1857+026, which is a larger and brighter source,  although both sources are considered distinct objects because  no significant emission connects them. The spectrum of HESS~J1858+020, over the range  0.5--80 TeV, follows a power law of spectral index $2.17\pm0.12$ and its flux is $3.5 \times 10^{-12}$ erg cm$^{-2}$ s$^{-1}$ \citep{aharonian08}. MAGIC (E $>$ 100 GeV) observed HESS~J1857+026 in 2010. In these observations, HESS~J1858+020 was also detected although with low significance 
given the relatively low MAGIC exposure at 0.5$^{\circ}$ angular distance  to the pointing positions \citep{stamescu12}. At GeV energies, 2FGL~J1857.6+0211 is positionally the closest {\it Fermi}-LAT source to HESS~J1858+020 that appears in the {\it Fermi} Large Area Telescope Second Source Catalogue \citep{2012ApJS..199...31N}. It is placed at 10$\arcmin$ from the HESS~J1858+020 location, with an error box radius for the source position
of about 4$\arcmin$, an energy flux of $\approx 10^{-10}$ erg cm$^{-2}$ s$^{-1}$ from 0.1 to 100 GeV, and a power-law index of 2.16.

At lower energies, this region was explored in the VLA Galactic Plane Survey (VGPS) at 1.4 GHz \citep{stil06} with 1$\arcmin$ resolution.  Using these data,  \cite{green09} identified the source  G35.6$-$0.4 as a supernova remnant, which is placed in the north-west border of HESS~J1858+020. 

More recently, \cite{paron11} used the $^{13}$CO (J=1-0) line from the Galactic Ring Survey \citep{2006ApJS..163..145J} to study the existence of molecular clouds towards the region of the supernova remnant (SNR) G35.6$-$0.4. They found 
a molecular cloud composed of two clumps, extending near the HESS source central region.
The study of the southeastern clump revealed the existence of a young stellar object, but the lack of evidence of molecular outflows, which would represent a jet able to produce gamma rays \citep{2007A&A...476.1289A}, seemed to discard the possibility of such an object as responsible of the observed TeV gamma-ray emission. It has also been proposed that  this emission is produced by hadronic interactions between the molecular cloud atoms and protons accelerated by the shock front of the SNR G35.6$-$0.4 \citep{paron11, 2003PhR...382..303T}.

To identify the (possible) radio counterpart of HESS~J1858+020,
we conducted exploratory observations with the Giant Metrewave Radio Telescope (GMRT) at 1.4~GHz and 610~MHz frequencies. 
These observations provide us with deep radio maps of the field of 
HESS~J1858+020 with arcsecond detail and improve the rms of the currently available images of the NVSS \citep{condon99} by one order of magnitude. As the data were taken at two different epochs, we can search for radio variability that is expected to be found in compact gamma-ray emitters such as microquasars and blazars.

In this work, we interpret the radio data from the region of the high energy source HESS~J1858+020 in a multi-wavelength context, in which the available IR, X-ray, and gamma-ray data are also included. 
The paper is structured as follows. In Sects.~\ref{observations} and \ref{arch}, we describe the radio observations, the data reduction, and the multi-wavelength archival data used; in Sects.~\ref{results-radio} and \ref{results-multi}, we present our results, and we discuss them in Sect.~\ref{discussion}.


\section{The GMRT observations}\label{observations}

We observed the HESS J1858+020 region at 610 MHz (49 cm) and 1.4 GHz
(21 cm) using the GMRT, located in
Pune (India). The radio observatory GMRT is an array of 30 antennas of 45 m diameter each
spread over distances of up to 25 km. We carried out the observations 
in three different epochs: May, June and July 2009 (see Table~\ref{gmrtlog}).  The angular size of
our target TeV source is about 10$\arcmin$, which is well within the primary
beam (FWHM) of GMRT at both 1.4 GHz (24$\arcmin$) and 610 MHz (43$\arcmin$). The
total observing time devoted to each frequency is detailed in Table \ref{gmrtlog}.
Observations were made with two 16-MHz sidebands (USB and LSB) centred
at 610 MHz and 1.4 GHz. The USB and LSB data were edited separately in AIPS{\footnote {The NRAO Astronomical Image Processing System.
http://www.aips.nrao.edu/}}.
The LSB data
were of relatively poor quality at 610 MHz, hence not included in the final images.
The flux density scale was set using 3C286 and 3C48 as primary amplitude
calibrators, and the phase calibration was performed by observing
repeatedly the phase calibrator 1822$-$096 at both frequencies.  A few
iterations of phase-only self-calibration and one iteration of amplitude
and phase self-calibration were performed to improve the image quality.
We have produced different maps for each frequency and run (day).

To detect variability, we observed at two different epochs at 610 MHz, which
allowed us to identify any possible variable radio source within the
HESS J1858+020 region observed. By averaging the two runs at 610 MHz, we
also produced a deeper radio image at this frequency.

\begin{table}
\caption[]{Log of the GMRT observations of HESS~J1858+020.}
\label{gmrtlog} 
\begin{tabular}{l@{\hspace{0.15cm}}c@{\hspace{0.15cm}}ccc@{\hspace{0.15cm}}c}
\hline
\hline
Date    &  Frequency &  Beam  size                                    &  PA             &  rms noise          & ON \\
 2009  &    [MHz]      &    [$\arcsec$] $\times$  [$\arcsec$] &  [$^{\circ}$]  &  mJy beam$^{-1}$         & source \\ 
          &                   &                                                     &                    &                          &  [hr] \\
 \hline 
 May 31 & 1400  &    $3.3 \times 3.0$      & 27   &    0.049     &   5.0   \\
 June 4  & 610    &   $16.8 \times 5.1$   & 43  & 0.26 & 5.0 \\
 July 8   & 610    &    $10.9 \times 4.5$  & 56  & 0.24  & 4.5 \\
  June 4 + July 8   & 610    &    $12.2 \times 4.8$  & 52  & 0.22  & 9.5 \\            
\hline
\end{tabular}
\end{table}
 
\section{Multi-wavelength archival data}\label{arch}
  
We explored the IR band by making use of data products at 3.4, 4.6, 12, and 22 $\mu$m from the Wide-field Infrared Survey Explorer (WISE) \citep{2010AJ....140.1868W}. 
We explored the X-ray band by analysing archival  {\it Chandra} data covering the central region of HESS~J1858+020. The X-ray data correspond to the project ObsID9116, and consist of 29.8 ks exposure observations carried out with the detector ACIS-I in mode VFAINT on June 21, 2008. The high energy gamma-ray data were obtained from the {\it Fermi}-LAT Second Source Catalogue (2FGL) \citep{2012ApJS..199...31N}.
  The information about the molecular cloud distribution in the region was obtained from the Galactic Ring Survey \citep{2006ApJS..163..145J}.
 
  \section{Radio sources around HESS~J1858+020}\label{results-radio}
  \subsection{GMRT results}
  
\begin{figure}
   \includegraphics[clip,angle=0,width=\columnwidth]{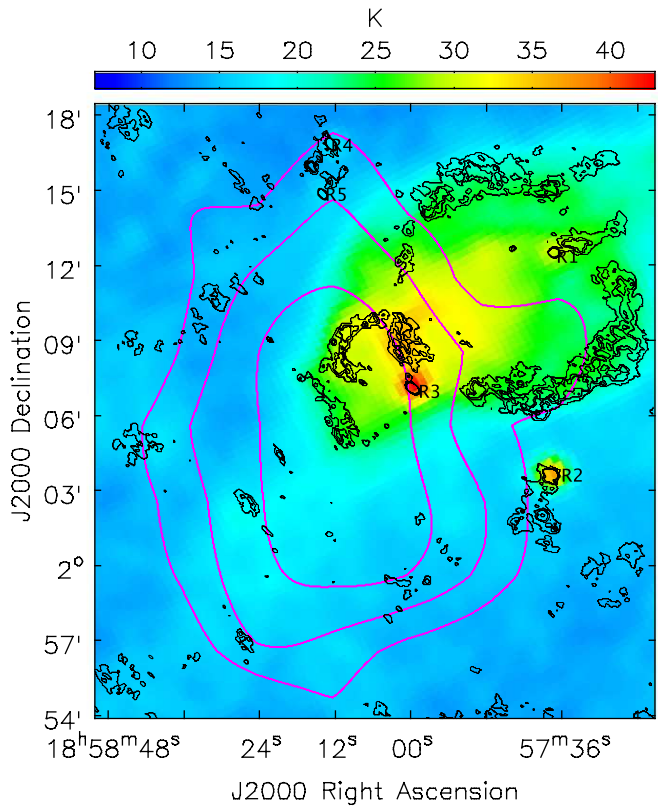}
      \caption{The
GMRT 610 MHz image is represented by black contours.  The contours correspond to 2, 3, and 4 times 0.22 mJy $\mathrm{beam^{-1}}$, the rms, and the beam size is $12.2\arcsec \times 4.8\arcsec$. The colour image shows the VGPS data of the supernova remnant G35.6$-$0.4 at 1.4 GHz with arc minute resolution \citep{stil06}. The HESS significance  contours (4, 5, and 6$\sigma$, in magenta) for the location of  HESS~J1858+020 are overlaid.
}
         \label{SNR}
   \end{figure}

  Figure~\ref{SNR} shows the two epochs average 610 MHz GMRT radio image (black contours)  of the field of HESS~J1858+020 with several arc second resolution.  This 610 MHz GMRT image is the most sensitive (few mJy beam$^{-1}$) and 
the highest resolution image ever obtained of this region. The colour image shows the VGPS data of this region at 1.4 GHz with arc minute resolution \citep{stil06}. The 1.4 GHz GMRT image does not show any relevant structures and, therefore, the image is not shown.

  The largest ring-shaped structure visible in the GMRT image of Fig.~\ref{SNR} follows the morphology of the SNR G35.6$-$0.4 \citep{green09}.   There is  also a smaller ring-shaped structure at the south-east end of the SNR, appearing as an independent structure. This structure, not resolved in previous radio images,  corresponds to an \ion{H}{ii} region that in the past led to consider G35.6$-$0.4 an \ion{H}{ii} region \citep{1989ApJS...71..469L}.  
Other relevant, compact, or slightly extended radio sources of this region detected above 5$\sigma$ are labelled with a number and are listed in Table~\ref{1858sources}.  
None of these five sources shows significant variability within uncertainties when comparing the flux density between June 4 and July 8.
 The most relevant sources are described  in more detail in what follows. 
   
  \subsection{Supernova remnant G35.6$-$0.4}
    
  The extended radio source G35.6$-$0.4, previously thought to be an \ion{H}{ii} region, was recently identified as a SNR \citep{green09}. 
  Its morphology was revealed by the 1 arcmin resolution images from the VGPS at 1.4 GHz \citep{stil06} (see Fig.~\ref{SNR}), and its non-thermal radio spectrum (spectral index  $\alpha=-0.47\pm0.07$) was obtained using data from surveys with low resolution (7.8 Jy at 1.4 GHz and of 3.1 Jy at 10 GHz) \citep{green09}. Observations of the \ion{H}{i} line in the VGPS provided a lower limit to the distance of about 3.7 kpc \citep{green09}. A recent kinematic distance study of the complex where the SNR G35.6$-$0.4 is placed has established a distance to the SNR of $3.6\pm0.4$ kpc \citep{2013arXiv1307.7878Z}.
  
\begin{figure*}
   \includegraphics[clip,angle=0,width=\textwidth]{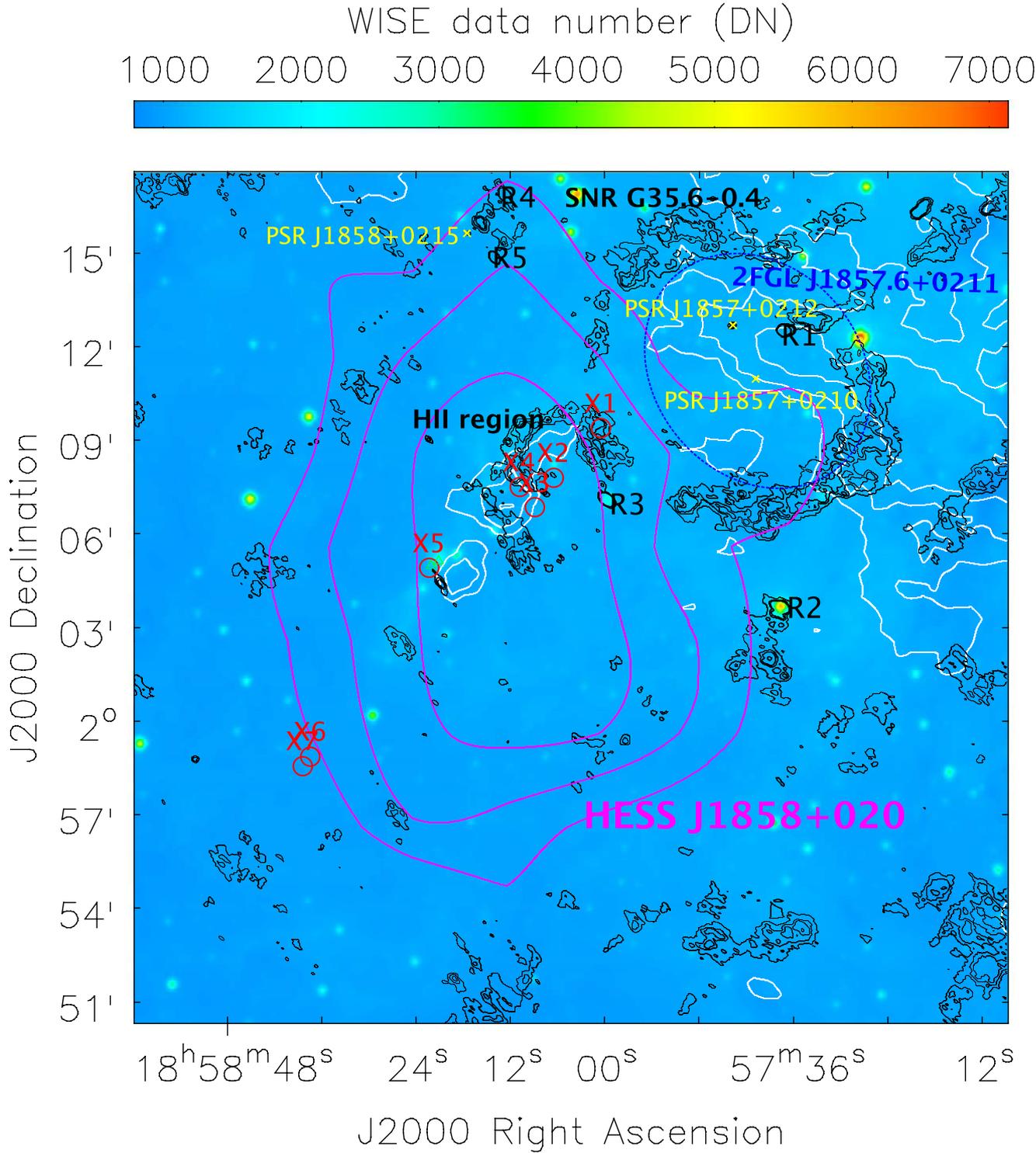}
      \caption{
      Radio image (black contours) of the field containing the unidentified H.E.S.S.
source HESS~J1858+020 obtained with GMRT at 610 MHz.  The contours correspond to 2, 3, and 4 times 0.22 mJy $\mathrm{beam^{-1}}$, the rms, and the beam size is $12.2\arcsec \times 4.8\arcsec$. The colour image shows the 12 $\mu$m WISE image of this region \citep{2010AJ....140.1868W}.
The white contours show the $^{13}$CO (J=1-0) emission integrated between 51 and 59  km s$^{-1}$ from the Galactic Ring Survey \citep{2006ApJS..163..145J}. The red circles correspond to X-ray sources. 
Some relevant radio sources 
are labelled as R1,..,R5, respectively (see Table~\ref{1858sources}). The shell-like structure clearly visible in the upper right region of the figure corresponds to the SNR G35.6$-$0.4.
The HESS significance  contours (4, 5, and 6$\sigma$, in magenta) for the location of  HESS~J1858+020 are overlaid. The blue ellipse indicates the 95$\%$ confidence location of the  {\it Fermi}-LAT gamma-ray source 2FGL~J1857.6+0211. The positions of the three pulsars are indicated with yellow crosses, and the positions of two young stellar objects, close to X5, are indicated by triangles.}
         \label{H1858_600}
   \end{figure*}

 Our image at 1.4 GHz does not show the SNR structure because our high spatial resolution resolves it out.
 Our 610 MHz image of the field containing HESS~J1858+020 shown in Figs.~\ref{SNR} and \ref{H1858_600}
 provides the highest resolution radio image to date (5--10 arc second) of SNR G35.6$-$0.4,  
 showing a nearly circular morphology ($10\arcmin \times 8\arcmin$ or $\approx 10\times8$ pc) that corresponds to the limb brightened region of emission. The integrated flux density of SNR G35.6$-$0.4 from our 610 MHz image is 0.46 Jy. 
 
   \begin{table*}
     \center
      \caption[]{Radio sources detected in the HESS~J1858+020 field with GMRT at 610 MHz and 1.4 GHz. We obtained the J2000.0 coordinates from the 1.4 GHz map. The 610 MHz flux density measurements were done in the combined image from June  4 and July 8 data. The spectral index $\alpha$ is defined as $S_{\nu} \propto \nu^{\alpha}$, where $S_{\nu}$ is the flux density at frequency $\nu$. The first column provides the assigned number to each source, ordered by increasing right ascension. The second and third columns list the J2000.0 right ascension and declination, respectively. The fourth and fifth columns provide the integrated flux density and its error at 610 MHz and 1.4 GHz, respectively. 
The last column lists the structure of the sources.} 
         \label{1858sources}
         \begin{tabular}{lcccccc}
            \hline
            \noalign{\smallskip}
            Source      &  $\alpha_{\rm J2000.0}$    &  $\delta_{\rm J2000.0}$  &  $S$$_{\rm 610\,MHz}$ &   $S$$_{\rm 1.4\,GHz}$ & Spectral     & Structure   \\
                               & [h\,m\,s]  & [$^{\circ}~\arcmin~\arcsec$]    &  [mJy]      & [mJy] &index $\alpha$ & \\ 
        
            \noalign{\smallskip}
            \hline
            \noalign{\smallskip}
                        R1   &     18 57 37.117  $\pm$    0.003  &  02 12 32.33  $\pm$     0.05    & 24.3 $\pm$ 1.1 & 24.0 $\pm$ 0.6  &$-0.02\pm$0.07 & \\   
                        R2   &     18 57 37.955  $\pm$    0.002  &  02 03 40.49  $\pm$     0.03    & 46.7 $\pm$ 1.2 & 118.5$\pm$ 2.0  &1.26$\pm$0.04 & Planetary Nebula \\                      
                        R3   &     18 57 59.578  $\pm$    0.007  &  02 07 06.97  $\pm$     0.10    & 50.6 $\pm$ 1.0 & 90.3 $\pm$ 6.3  & 0.78$\pm$0.10&Planetary Nebula\\ 
                        R4   &     18 58 12.557  $\pm$    0.003  &  02 16 52.94  $\pm$     0.04    & 31.0 $\pm$ 0.4 & 24.8 $\pm$ 0.5   &$-0.30\pm$0.03 & \\  
                        R5   &     18 58 13.898  $\pm$    0.004  &  02 14 52.05  $\pm$     0.05    & 21.7 $\pm$ 0.2 & 13.1 $\pm$ 0.2   &$-0.68\pm$0.02 & \\            
            \noalign{\smallskip}
            \hline
         \end{tabular}
         \end{table*}

 Within the field of the SNR, there are two known pulsars that we have not detected. The pulsar PSR J1857+0212 (PSR B1855+02), with a period of 0.417 seconds, was proposed to be associated with the remnant G35.6$-$0.4 \citep{1993ASPC...35..419P}. The characteristic age is $\sim 1.6\times 10^5$~yr \citep{2004MNRAS.353.1311H} and the estimated distance, derived from the observed dispersion measure, is $\sim$ 8 kpc \citep{2005MNRAS.360..974H}. The other pulsar is PSR J1857+0210, which is close to the centre of the SNR, and has a flux density at 1.4 GHz of 0.30(6) mJy, a period of 0.63 s, a characteristic age of $7\times 10^{5}$~yr, and an estimated distance of 15.4 kpc \citep{2002MNRAS.335..275M}. The expected age of the SNR, deduced from a Sedov-Taylor model, is about $3\times 10^{4}$~yr,  which poses difficulties for the association of the SNR with any of these pulsars \citep{green09}. However, the not well established link between the characteristic and the true age of the pulsars prevents us from ruling out this association.

\subsection{The \ion{H}{ii} region}

Radio recombination lines towards $l=35^{\circ}.588$, $b=-0^{\circ}.489$ were detected in a survey of radio \ion{H}{ii}  regions in the northern sky \citep{1989ApJS...71..469L}. Further detections of recombination lines (H168$\alpha$ at 1374.6006 MHz) in the same region show broad lines indicative of enhanced turbulence caused by a non-thermal source  \citep{1993ASPC...35..419P}. This source could be the SNR G35.6$-$0.4 and this would imply that the \ion{H}{ii}  is located at the same distance from the SNR, i.e., $3.6\pm0.4$ kpc \citep{2013arXiv1307.7878Z}. From our results shown in Fig.~\ref{H1858_600}, we can see that the \ion{H}{ii}  region is positionally coincident ($\alpha_{\rm J2000.0}$ =18h\,58m and $\delta_{\rm J2000.0}$ = 2$^{\circ}~$8$\arcmin$) with the ring-shaped radio structure that appears in the southern corner of the SNR G35.6$-$0.4. The size  of the ring is about 3$\arcmin$, which corresponds to a linear size of 3 pc adopting the distance of 3.6 kpc.
This structure seems to be morphologically independent of the SNR, although there could be some physical influence (turbulence) from the SNR. The \ion{H}{ii} region is nearly positionally coincident with the centroid of HESS~J1858+020. 
   \begin{figure*}
 \includegraphics[clip,angle=0,width=\textwidth]{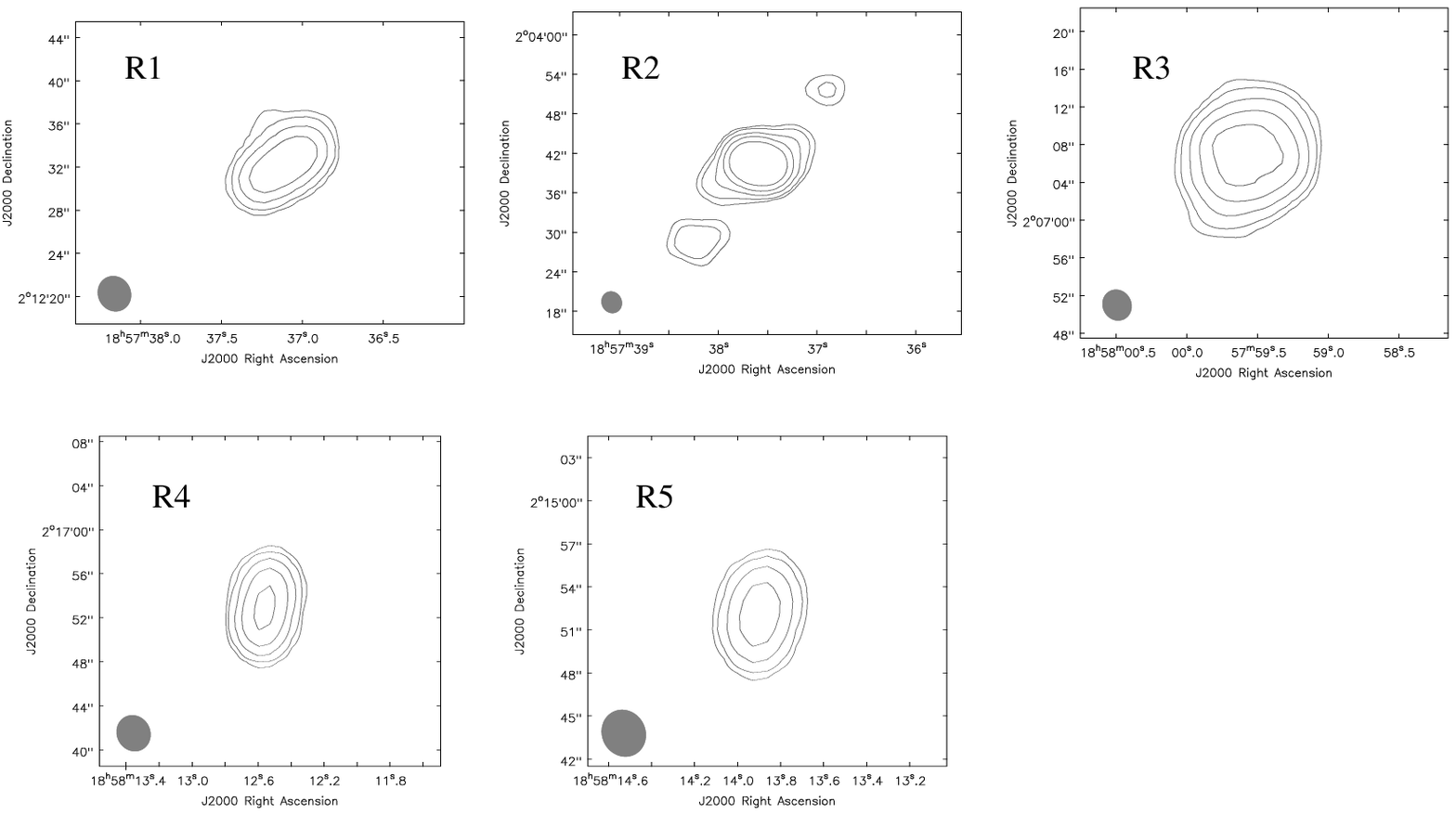}
      \caption{The GMRT image at 1.4 GHz of the five sources listed  in Table~\ref{1858sources}. The radio contours correspond to 5, 8, 16, 32, and 64 times 0.049 mJy beam$^{-1}$, the rms noise. The synthesized radio beam is plotted in the lower-left corner. Sources R2 and R3 are catalogued as possible and confirmed planetary nebula respectively.}
         \label{u}
   \end{figure*}

\subsection{Other sources}



Source R1: this source is clearly detected by GMRT and resolved at both frequencies, showing an elongated morphology of about 10 arc seconds in length at 1.4 GHz (see Fig.~\ref{u}-top left) and a more symmetric morphology of  about 25 arc seconds at 610 MHz. The flux density at 610 MHz is 24.3$\pm$1.1 mJy and at 1.4 GHz is 25.2$\pm$  0.7 mJy, implying a flat spectral index. 
This source also appears in the Small-Diameter Radio Sources Catalogue \citep{zoonematkermani90} with an integrated flux density of  21 mJy at 1.4 GHz, and in the Compact Radio Sources in the Galactic Plane Catalogue \citep{2005AJ....130..586W}, with an integrated flux of 10.45 mJy at 5 GHz. There is no counterpart in the 2MASS archive data. The source is outside the significance contours of HESS~J1858+020, but within the {\it Fermi}-LAT source 2FGL~J1857.6+0211 position error box (see Fig.~\ref{H1858_600}). 

Source R2: this is one of the strongest sources in the field. Its location, well outside the significance contours of HESS~J1858+020 and the error box of 2FGL~J1857.6+0211, makes this source an unlikely counterpart of these gamma-ray sources.  Detected at both frequencies,  46.7 $\pm$1.2 mJy at 610 MHz and 125$\pm$4 mJy at 1.4 GHz, source R2 shows an elongated structure 
(see the top middle panel of Fig.~\ref{u}). Our 1.4 GHz flux density is similar to the value of 117$\pm$4 mJy obtained by \cite{condon99}. \cite{1995AJ....110.2225K} catalogued the source 2MASS J18573787+0203440 (IRAS 18551+0159), positionally coincident with source R2, as a planetary nebula. \cite{2013arXiv1307.7878Z} estimated a distance of $4.3\pm0.5$ kpc through kinematic distance studies.

Source R3: this radio source is positionally coincident with IRAS 18554+0203. It has been detected clearly at both frequencies with the flux density at 1.4 GHz nearly doubling the value obtained at 600 MHz. At both frequencies it has been resolved, showing a round structure  of $\approx 12\arcsec$ at 1.4 GHz (see the top right panel of Fig.~\ref{u}) and $\approx 30\arcsec$ at 600 MHz. This source has been catalogued as a Galactic planetary nebula (PN G035.5$-$00.4,  PHR 1857+0207), showing a radio flux density of 100$\pm$4 mJy at 843 MHz  and an optically determined angular diameter in MASH of $11\arcsec$  \citep{2011MNRAS.412..223B}. \cite{2013arXiv1307.7878Z} estimated a lower limit distance of 3.8$\pm$0.4 kpc and an upper limit distance of $5.4\pm0.7$ kpc through kinematic distance studies.


Source R4: this source has been resolved with GMRT at 1.4 GHz, showing a nearly north-south elongated structure (see the bottom left panel of Fig.~\ref{u}). The source appears in the 20 cm source catalogue of compact radio sources in the Galactic plane \citep{2005AJ....130..586W} with an integrated flux density of 24 mJy and with hints of north-south elongation and with a flux density of 23$\pm$0.3 mJy in the 5 GHz CORNISH catalogue \citep{2013ApJS..205....1P}. There are two optical/infrared sources, 2MASS 18581274+0216523 and 2MASS 18581239+0216513, placed within the radio contours. Our measured spectral index (see Table 2) suggests a non-thermal nature.

Source R5: this source appears in the 20 cm source catalogue of compact radio sources in the Galactic plane with an integrated flux density of 6.7 mJy, whereas it appears resolved in our GMRT 1.4 GHz observations (see the bottom middle panel of Fig.~\ref{u}). In the CORNISH catalogue, this source has a flux density of 4.3$\pm$0.4 mJy at 5 GHz. Our measured spectral index (see Table 2) suggests a non-thermal nature.

PSR J1858+0215: although we have not detected this object, we note the existence of this source, which is located in the northern part of HESS J1858+020, within its 4 and 5$\sigma$ confidence contour levels and very close to sources R4 and R5. Its flux density at 1.4 GHz is 0.22(4) mJy, the period is 0.75 s, the age is of about $2.6\times 10^{6}$ years, and the distance is12.4 kpc \citep{2002MNRAS.335..275M}.

\section{Multi-wavelength study}\label{results-multi}

The results obtained from the analysis of archival  data and the data from different surveys, as well as our GMRT radio data, are shown in Fig.~\ref{H1858_600}. 

\subsection{Molecular cloud}

\cite{paron10} and \cite{paron11} revealed the existence of a molecular cloud, formed by two clumps, in the central region of HESS~J1858+020, using the $^{13}$CO (J=1-0) line from the Galactic Ring Survey. The northern clump shows kinematical hints of shocked gas that could evidence turbulent motion of the gas. This turbulent motion was suggested  to be caused by the interaction of the SNR G35.6$-$0.4 with the molecular cloud \citep{paron10}. In Fig.~\ref{H1858_600}, we can see that the ring-shaped structure detected in our 610 MHz observations overlaps with the northern clump, suggesting an association between the {\ion{H}{ii}} region and the molecular cloud. In fact, the study of the impact of stellar winds, supernovae, and {\ion{H}{ii}} regions onto cloud turbulence shows that {\ion{H}{ii}} regions most efficiently sustain this turbulence \citep{2002ApJ...566..302M}. On other hand, kinematic absorption features shown in \cite{2013arXiv1307.7878Z} make a common distance for both the molecular cloud and the supernova plausible.

The southern molecular clump has no radio counterpart, as can be seen in Fig.~\ref{H1858_600}. However, \cite{paron11} identified two intrinsically  red sources in the infrared domain, named IRS1 ($\alpha_{\rm J2000.0}$ =18h\,58m\,21.15s and $\delta_{\rm J2000.0}$ = 2$^{\circ}~$5$\arcmin~$10.0$\arcsec$) and IRS2 ($\alpha_{\rm J2000.0}$ =18h\,58m\,22.12s and $\delta_{\rm J2000.0}$ = 2$^{\circ}~$5$\arcmin~$1.4$\arcsec$), which appear to be related to the molecular cloud. These authors suggest that IRS1 is an evolved young stellar object, probably in the last stages of formation, which could be  a high-mass protostellar object embedded in the southern molecular clump. It appears as a bright point source (mag 6.51$\pm$0.03) in the 12$\mu$m image.

\subsection{{\it Chandra} X-ray sources}

In the 0.4--7 keV image taken with {\it Chandra} around the HESS source,
seven X-ray sources are detected by {\it celldetect} at significance
above the conservative $3\sigma $ threshold. The locations of these
sources are indicated in Fig.~\ref{H1858_600} by red circles. Table~\ref{xray} lists the
position, 0.4--7 keV net counts, the significance of the {\it celldetect}
detection, and the hardness ratio defined as HR = (H$-$S)/(H$+$S) 
(where H and S are source counts in the 3--7 keV and 1--3 keV bands,
respectively). The HR of X1,2,3,4 all have positive values, implying
significantly harder spectra than X5 and X6. None of the seven sources
has an IR counterpart. Four of the detected sources, X1,2,3,4,
coincide with the northern molecular clump and the ring-shaped radio
structure. 

\begin{figure}
    \includegraphics[clip,angle=90,width=\columnwidth]{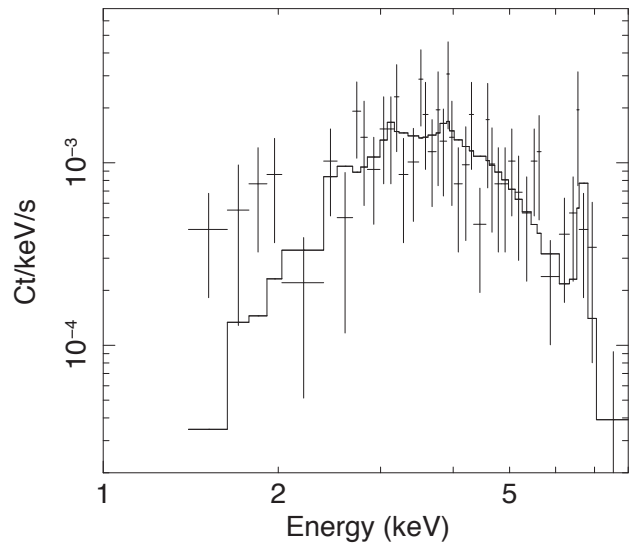}
      \caption{The X-ray spectrum of the summed X1-4 sources, obtained from Chandra ACIS-I. No X-ray emission below 1 keV is detected. The solid-line histogram shows the folded model of a thermal emission spectrum of $kT = 2$ keV modified by absorption of $N_{\rm H}=1.1\times 10^{23}$ cm$^{-2}$. }
         \label{sum}
   \end{figure}

As their HR suggest, X1-4 have similar spectra and the combined (a
straight sum) spectrum of the four sources (Fig.~\ref{sum}) shows strong
absorption and a hint of an Fe K emission line at $6.53\pm 0.09$ keV
(at $\sim 1.5\sigma $). If this Fe K line were real, it would
suggest a thermal origin of the X-ray emission. Whereas the temperature
and absorption are degenerate in spectral fit and are not well
constrained, when solar abundances are assumed, the likely temperature of
the thermal emitting gas is $kT\sim $1.5--2 keV (which provides a good
description of the Fe K line energy and strength) with an absorbing
column density of $N_{\rm H}\sim $(1--1.5)$\times 10^{23}$
cm$^{-2}$. These spectral properties suggest that these sources might be
embedded protostars. The mean flux of these sources as observed in the
2--7 keV band is $2.6\times 10^{-14}$ erg s$^{-1}$ cm$^{-2}$.

\begin{figure}
    \includegraphics[clip,angle=90,width=\columnwidth]{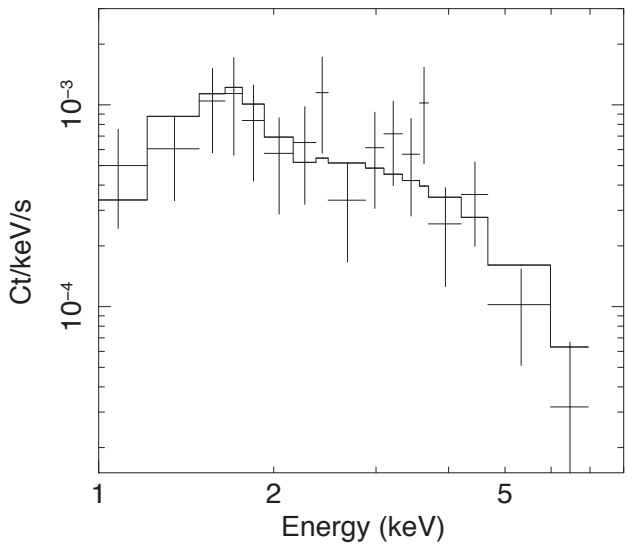}
      \caption{The X-ray spectrum of X5. No significant X-ray emission is detected below 1 keV. The solid-line histogram indicates the folded model of a power-law of $\Gamma = 2.1$ modified by absorption with $N_{\rm H}=1.5\times 10^{22}$ cm$^{-2}$. }
         \label{c5}
   \end{figure}

The brightest source in the 0.4--7 keV band, X5, is weaker than
X1--4 in the 3--7 keV band. A power-law of index $\Gamma=2.1\pm 0.6$ and a
thermal spectrum with $kT = 3.2\pm 1.1$ keV yield comparable fits to
the data (Fig.~\ref{c5}). The absorption $N_{\rm H}=(1.5\pm 0.6)\times
10^{22}$ cm$^{-2}$ agrees with the typical Galactic absorption towards this
direction of the sky for a distance of several kpc. The observed 2--7 keV flux is $2.7\times 10^{-14}$
erg s$^{-1}$ cm$^{-2}$. This source is potentially coincident with the star
forming region and close to the southern molecular clump.

The source X6 has the softest spectrum and, unlike other sources, significant emission is detected below
1 keV indicating low absorption. This probably means that this is a foreground source. 
The faintest of the X-ray sources detected is X7, and it is not possible to get a reliable spectrum.

Besides the point-like sources, diffuse emission, possibly associated
with HESS~J1858+020, was searched in the Chandra data. As such
diffuse emission could have an extension on arcmin-scale, the X-ray
image coarsely binned with various angular scales was inspected, but no
clear extended emission was found. Based on the statistical
fluctuation of the background of the field, the $2\sigma $ upper limit
was obtained to be $\sim 1\times 10^{-13}$ erg cm$^{-2}$ s$^{-1}$ for
a source with a $6^{\prime}\times 6^{\prime }$ extension and an X-ray
spectrum of a power-law of photon-index $\Gamma = 2$ (the flux was
corrected for the Galactic absorption, approximated to be $N_{\rm
H} = 1.5\times 10^{22}$ cm$^{-2}$).

   \begin{table}
      \caption[]{X-ray sources detected after the analysis of archival {\it Chandra} data covering the central region of HESS~J1858+020.}
      \label{xray}
      \begin{tabular}{l@{\hspace{0.10cm}}c@{\hspace{0.15cm}}c@{\hspace{0.15cm}}c@{\hspace{0.15cm}}c@{\hspace{0.03cm}}r}
            \hline
            \noalign{\smallskip}
            Source      &  $\alpha_{\rm J2000.0}$    &  $\delta_{\rm J2000.0}$         &  Counts  &   Significance  &  HR~~~~~  \\
                             & [h\,m\,s]                         & [$^{\circ}~\arcmin~\arcsec$]  &  [0.4--7\,keV]    &           &     \\
        
            \noalign{\smallskip}
            \hline
            \noalign{\smallskip}
                        X1   &    18 58 00.46    &   02 09 23.3  & 35.04 $\pm$8.27 & 4.23 &  0.38 $\pm$0.18 \\                      
                        X2   &    18 58 06.49    &   02 07 47.7  & 22.43 $\pm$6.97 & 3.22 &  0.72 $\pm$0.24 \\   
                        X3   &    18 58 08.97    &   02 06 51.2  & 22.24 $\pm$6.65 & 3.34 &  0.55 $\pm$0.23 \\ 
                        X4   &    18 58 10.71    &   02 07 30.1  & 40.14 $\pm$8.57 & 4.69 &  0.45 $\pm$0.16 \\  
                        X5   &    18 58 22.21    &   02 04 54.6  & 64.43 $\pm$9.94 & 6.55 &  $-$0.13 $\pm$0.12 \\       
                        X6   &    18 58 37.38    &   01 58 51.5  & 62.24 $\pm$9.50 & 6.55 &  $-$0.80 $\pm$0.20 \\  
                        X7   &    18 58 38.50    &   01 58 33.2  & 20.69 $\pm$6.80 & 3.04 &  0.05 $\pm$0.22 \\       
            \noalign{\smallskip}
            \hline
         \end{tabular}
         \end{table}

\subsection{The high energy gamma-ray source 2FGL~J1857.6+0211}
The source 2FGL~J1857.6+0211 is positionally the closest {\it Fermi}-LAT source to HESS~J1858+020 that appears in the {\it Fermi} Large Area Telescope Second Source Catalogue \citep{2012ApJS..199...31N}. However, a likelihood analysis performed to investigate a possible association between  HESS~J1858+020 and 2FGL~J1857.6+0211 \citep{2011MNRAS.417.3072T} suggests that this association is unlikely (see however Sect.~\ref{discussion}). The {\it Fermi}-LAT source, with a semi-major and semi-minor axis of position error ellipse at 95\% confidence of $4\arcmin\times3\arcmin$, remains unidentified. 

The pulsars PSR J1857+0212 and PSR J1857+0210 are both located well inside the 2FGL~J1857.6+0211 error box. {\it Fermi}-LAT  has not yet detected pulsed gamma-ray emission from any of the pulsars.
  
  \section{Discussion}\label{discussion}
  
It has been proposed that the TeV $\gamma$-ray emission detected by H.E.S.S. has a hadronic origin and is produced by the
interaction between cosmic rays accelerated by the shock front of the SNR and a nearby molecular cloud \citep{paron11}.
\cite{2011MNRAS.417.3072T} explored this possibility through the study of the parameter space considering different cloud
masses, cloud-SNR distances, and spatial diffusion coefficients. They concluded that only a rather small diffusion coefficient,
$\sim 100$ times lower than the Galactic one, and a short distance from the SNR, of about 40~pc, could reconcile the detection
of TeV emission, the GeV upper limits at the location of the HESS source, and the cloud mass derived by \cite{paron10}.
Nevertheless, given the small projected distance of $\sim 5$ pc (at a distance of 3.6 kpc) between the cloud and the SNR, plus the fact that close to SNR shocks particle
propagation may be significantly slowed down \citep{fujita11,malkov13}, the connection between the TeV emission and the SNR
cannot be discarded. In fact, new calculations modelling the interaction between the 
SNR cosmic rays and the cloud, using the distance of 3.6~kpc, relax the conditions on the diffusion coefficient and cloud mass, 
making this possibility more likely than previously thought \citep[see discussion in][]{2013arXiv1307.7878Z}. The X-ray upper limit for the SNR G35.6$-$0.4 and MC interaction region we found can help us to constrain, in future modelling, the physical conditions of the TeV emitter, like the magnetic field and the medium density.

The age of the SNR, probably close to its transition to the radiative phase, but possibly still under adiabatic expansion
\citep{2011MNRAS.417.3072T}, suggests that it could still accelerate protons energetic enough to produce GeV gamma rays. 
The recently estimated distance of 3.6~kpc makes the energetics of the source less tight and favors this possibility.
The SNR origin could explain the co-spatial {\it Fermi}-LAT source 2FGL~J1857.6+0211, as in other middle-aged SNR, through interactions
with very nearby clouds or shock-embedded dense material \citep[see, e.g.,][and references therein]{ajello12}. 
Therefore, it seems feasible that SNR G35.6$-$0.4 would be the origin of both the GeV and TeV emission.

Other sources presented in this work may be behind HESS~J1858+020 and/or 2FGL~J1857.6+0211, although their association seems to
be less plausible than SNR G35.6$-$0.4. There are the three pulsars discussed above. Being rather old and far, with
distances $\gtrsim 10$~kpc, their spin-down luminosities are certainly not enough to power the GeV emission centred at the
location of the SNR, and may be insufficient to explain the TeV emission. 
In addition, all three pulsars lie outside the
significance contours of the HESS source, making their connection with the TeV radiation even more unlikely. 
However, one of these objects may still be behind an ancient pulsar wind nebula, with the pulsar being off-set and having a complex morphology and no X-ray radiation \citep[see, e.g.,][]{2010AIPC.1248...25K}, although the age may argue against an association with 2FGL~J1857.6+0211 \citep{2013arXiv1305.2552K}. 
The two planetary
nebulae also present in the region can be discarded as counterparts of the gamma-ray sources as well. These objects, remnants of
low-to-intermediate mass stars, would have limited energetics and relatively low speeds, which are unsuitable conditions to accelerate
particles up to very high energies. The two non-thermal radio sources (R4 and R5), with unclear counterpart at other wavelengths,
might be candidates for high energy emitters, although at present assessing this statement requires further study to identify
their physical origin. Finally, the star forming region deserves some attention, in particular the elongated $^{13}$CO structure and several X-ray sources with positions roughly aligned with the former. As mentioned, this region could host a number of protostars, with at least one among them possibly of high mass nature. Moreover, the apparent elongation of the region may be hinting at a powerful outflow, which could be masked in radio by strong free-free absorption. Such an outflow, and the entire star formation region, could be the origin of, or contribute to,
HESS~J1858+020 \citep[e.g.][]{2010A&A...511A...8B}. However, this possibility requires further observational studies to sustain the case for strong interactions in the star formation region as well.

\section{Conclusions}

The region surrounding HESS~J1858+020 is complex. For instance, there are several molecular structures, possibly linked with different types of non-thermal objects, like SNR or star forming regions, and these are suitable environments for the formation of high energy emission. In this regard, several candidates can be identified as potential counterparts to the HESS (and possibly also the {\it Fermi}) source: a SNR interacting with very near molecular clouds through a medium with a low diffusion coefficient; a star forming region with several young stellar objects, some of them possibly of high mass; and non-thermal extended radio sources that may be Galactic in origin and host relativistic jets. Finally, an ancient pulsar wind nebula, despite the far distance of the pulsars neighbouring the HESS source, cannot be discarded as a candidate. In any case, given the natural uncertainties when determining the most likely candidate associated with the richness of the presented results, further observational studies are needed to narrow the search for a counterpart to HESS~J1858+020.

\begin{acknowledgements}
J.M.P., V.B-R, V.Z. and M.R. acknowledge support by DGI of the Spanish Ministerio de Econom\'{\i}a y Competitividad (MINECO) under grants AYA2010-21782-C03-01 and FPA2010-22056-C06-02. 
V.B-R. acknowledges financial support from MINECO through a Ram\'on y Cajal fellowship.
 J.M.P. acknowledge financial support from ICREA Academia. V.Z.~acknowledges support by the Generalitat de Catalunya through the
Beatriu de Pin\'os program, and by the Max-Planck-Gesellschaft. This research has been supported by the Marie Curie Career Integration
Grant 321520.
We thank the staff of GMRT who have made these observations
possible. GMRT is run by the National Centre for Radio Astrophysics
of the Tata Institute of Fundamental Research. This publication makes use of data products from the Wide-field Infrared Survey Explorer, which is a joint project of the University of California, Los Angeles, and the Jet Propulsion Laboratory/California Institute of Technology, funded by the National Aeronautics and Space Administration. This publication makes use of molecular line data from the Boston University-FCRAO Galactic Ring Survey (GRS). The GRS is a joint project of Boston University and Five College Radio Astronomy Observatory, funded by the National Science Foundation under grants AST-9800334, AST-0098562, AST-0100793, AST-0228993, \& AST-0507657.
\end{acknowledgements}

\bibliographystyle{aa}


\end{document}